\documentstyle[12pt]{article}
\fussy
\newcommand{\FF}{{\cal F}}

\renewcommand{\=}{&=&} 

\newcommand{\eg}{{{\em e.g.},\ }}
\renewcommand{\bar}{\overline}
\renewcommand{\a}{\alpha}
\renewcommand{\b}{\beta}

\newcommand{\g}{\gamma}

\newcommand{\m}{\mu}

\renewcommand{\o}{\omega}
\newcommand{\s}{\sigma}
\renewcommand{\t}{\theta}
\newcommand{\tb}{{\bar\theta}}
\newcommand{\aad}{{\a\ad}}
\newcommand{\ad}{{\dot{\alpha}}}
\newcommand{\bd}{{\dot{\beta}}}
\newcommand{\pa}{\partial}
\newcommand{\th}{\widehat\theta}
\newcommand{\yh}{\widehat{y}}
\newcommand{\ybh}{\widehat{\bar y}}
\newcommand{\tbh}{\widehat{\bar\theta}}

\newcommand{\gd}{{\dot{\gamma}}}
\newcommand{\qh}{{\widehat Q}}
\newcommand{\qbh}{{\widehat{\bar Q}}}
\newcommand{\dh}{{\widehat D}}
\newcommand{\dbh}{{\widehat{\bar D}}}

\newcommand{\bb}{{\beta}}

\newcommand{\xx}{{\bf x}}

\newcommand{\real}{{{\rm I} \kern -.19em {\rm R}}}

\renewcommand{\ss}{\sigma\kern-.54em \sigma}
\newcommand{\pab}{\overline{\pa}}
\newcommand{\pah}{\widehat\pa}
\newcommand{\pabh}{\widehat{\overline{\pa}}}
\newcommand{\be}{\begin{equation}}
\newcommand{\ee}{\end{equation}}
\newcommand{\beq}{\begin{eqnarray}}
\newcommand{\eeq}{\end{eqnarray}}
\newcommand{\ba}{\begin{array}}
\newcommand{\ea}{\end{array}}
\newcommand{\equ}[1]{(\ref{#1})}
\newcommand{\nn}{\nonumber}

\begin{document}
January 1999\hfill hep-th/9901052\\[-1mm]

\begin{center}
{\large\bf FINITE-TEMPERATURE SUPERSYMMETRY\\ AS A CONSTRAINED 
SUPERGRAVITY\footnote{Talk presented at 
``Strong and Electroweak Matter 98", 
Copenhagen, December 2-5 1998.}}\\[6mm]

C. LUCCHESI\footnote{Supported in part by the Swiss National 
Science Foundation.}\\[1mm]

Institut de Physique, 1 rue Breguet, 2000 Neuch\^atel (Switzerland)\\
E-mail: Claudio.Lucchesi@iph.unine.ch\\[4mm]


\begin{minipage}{12truecm}
                 \footnotesize
                 \parindent=0pt 
{We introduce thermal superspace and show how it 
can be used to reconcile the superfield formulation of supersymmetry 
with finite temperature environments.}\par
                 \end{minipage}
\end{center}
%

\vspace{4mm}
\noindent{\large\bf 1 \ Thermal Superspace}
\vspace{2mm}

%
Immersing a physical system in a heat bath results in the fields
acquiring different properties according to their statistics.
{\it E.g.}, finite-temperature bosonic fields obey periodic boundary
conditions, while fermionic fields satisfy antiperiodic b.c.'s.
Such a distinction can be seen also at the level of the Green's functions.
Depending on the field's statistics, thermal propagators obey either a
bosonic KMS condition, or a fermionic one. Therefore, thermal
effects induce a clear and mandatory distinction between bosons and
fermions. As a consequence, finite temperature environments are
incompatible with $T=0$ supersymmetry~: the supersymmetry transformation
is indeed unable to take into account the distinct thermal properties
that go along with different statistics.

The thermal superspace approach~\cite{dl} allows to formulate supersymmetry 
at finite temperature in a way which respects
the different thermal behaviours of bosons and fermions. 
The parameters of supersymmetry transformations at $T\!>\!0$ being 
{\it time-dependent} and {\it antiperiodic}~\cite{ggs} on the imaginary 
time interval $[0,\b]$, thermal supersymmetry takes 
periodic bosons into antiperiodic fermions, and {\it vice-versa}.
Assuming the corresponding $T\!>\!0$ supersymmetry charges to induce motion 
in some superspace,
we naturally require the Grassmann coordinates of that superspace to be
{\it time-dependent} and {\it antiperiodic}, like the thermal supersymmetry 
parameters.
A point in thermal superspace has therefore coordinates
$[x^\mu,{\th}^\a(t),{\tbh}^\ad(t)]$, $t=x^0$,
where a ``hat" is used to denote thermal quantities,
and the thermal Grassmann coordinates are subject to
the temperature-dependent {\it antiperiodicity conditions}
\vspace{-1.5mm}
\be
{\th}^\a(t+i\b)=-{\th}^\a(t)\, ,\qquad
{\tbh}^\ad(t+i\b)=-{\tbh}^\ad(t)\,,\quad {\rm with} \ \ \b={1\over T}\,.
\label{antiper}
\ee
So, thermal superspace -- in which we shall 
define thermal superfields, thermal supercharges, etc. (see below) --
reconciles supersymmetry with finite temperature.
This approach is particularly welcome for, \eg cosmology, 
as one can now contemplate 
formulating supergraph techniques at finite temperature, and deriving 
from these the $T>0$ effective potential in superfield form~\cite{bl}. 

Thermal supersymmetry corresponds to a transformation with 
time-de\-pen\-dent, that is, {\it local} parameters. It is in this sense 
a form of supergravity, but a restricted one, as locality is enforced 
only in the time direction, and as, moreover, the thermal supersymmetry 
parameters are subject to an antiper\-iodicity condition identical to 
equation \equ{antiper}.
%
%

\vspace{4mm}
\noindent{\large\bf 2 \ Thermal Superfields, Super-KMS Condition 
\& Super-B.C.}
\vspace{-2mm}

%
At $T\!=\!0$, supermultiplets are conveniently described in the 
language of superfields. Here we define {\it thermal} 
superfields as expansions in thermal superspace, and show 
how these can be used to write consistent boundary and KMS~\cite{kms} 
conditions. In taking $t$-dependent superspace Grassmanns, we 
introduce a new $t$-dependence in the chiral variables,  which we define 
at $T>0$ through
$\widehat y^\mu_{(t)}
=
x^\m -i\th(t)\s^\m\tbh(t)$.
{\it Thermal chiral superfields} can then be expanded as
\vspace{-1.5mm}
\be
\widehat\phi[\widehat y_{(t)},\th(t)]
=
z[\widehat y_{(t)}]+\sqrt{2}\,\th(t)\,\psi[\widehat y_{(t)}]
- \th(t)\th(t)\, f[\widehat y_{(t)}]\,,
\label{tch}
\ee
and thermal antichiral ones follow a similar expansion. The thermal 
(chiral-anti\-chi\-ral) superfield propagator\footnote{We write $\yh_i$, 
$\th_i$, $\tbh_i$ in place of
$\yh_i(t_i)$ and $\th_i(t_i)$, $\tbh_i(t_i)$.}
 $G_C\,[\widehat y_{1}, \widehat{\bar y}_{2},\th_1,
\tbh_2]\equiv
\langle {\sf T}_{\!C}\  \widehat\phi[\widehat y_{1},\th_1]
\,\widehat{\bar\phi}\,[\widehat{\bar y}_{2},\tbh_2] \rangle_\bb$ expands, 
in analogy to $T\!=\!0$, as
\vspace{-1.5mm}
$$
G_C[\widehat y_{1}, \widehat{\bar
y}_{2},\th_1,\tbh_2]
= D_C[\yh_{1}\! -\! \ybh_{2}]
-2\th_1^\a\tbh_{2\,\bd}S_{C\,
\a}^{\ \ \ \bd}[\widehat y_{1}\! -\! \widehat{\bar y}_{2}]
+ \th_1\th_1\,
\tbh_2 \tbh_2\,\FF_C [\widehat y_{1} -
\widehat{\bar y}_{2}]\,,
$$
where $D_C$, $S_C$ and $\FF_C$ denote the thermal propagators for the 
component fields $z$, $\psi$ and $f$, respectively.
Thermal field theory requires the propagators of thermal scalar components 
to obey a bosonic KMS condition, 
\vspace{-1.5mm}
\beq
D_{C}^>(t_1;\xx_1,t_2;\xx_2)
\= D_{C}^< (t_1+i\b;\xx_1,t_2;\xx_2)
\,,\label{c1}\\[1mm]
\FF_C ^>(t_1;\xx_1,t_2;\xx_2)
\= \FF_C^< (t_1+i\b;\xx_1,t_2;\xx_2)
\,,\label{c3}
\eeq
while the thermal fermionic component must fulfill a fermionic KMS constraint,
\vspace{-1.5mm}
\be
S_{C\,\a}^{>\ \ \bd} (t_1;\xx_1,t_2;\xx_2)
= -
S_{C\,\a}^{< \ \ \bd} (t_1+i\b;\xx_1,t_2;\xx_2)\,.
\label{c2}
\ee
Thermal superspace allows to write a KMS condition at the level of thermal 
{\it superfield} propagators. Indeed, defining 
$G_C^> =\langle
\widehat\phi\,\widehat{\bar\phi}\rangle_\bb$, 
$G_C^< =\langle \widehat{\bar\phi}\,\widehat\phi\rangle_\bb$,
a {\it superfield KMS (or super-KMS) condition} can be written as~:
\vspace{-1.5mm}
\be
G_C^>\,[\widehat y_{1(t_1)}, \widehat{\bar
y}_{2(t_2)},\th_1(t_1),\tbh_2(t_2)]
=
G_C^<\,[\widehat y_{1(t_1 + i\beta)}, \widehat{\bar
y}_{2(t_2)},\th_1(t_1+i\b),\tbh_2(t_2)]\,,
\label{skms}
\ee
where $\widehat y_{1(t_1 + i\beta)}=
\widehat y_{1(t_1)} + (i\beta\,\,;\,\,{\bf 0})$.
Clearly, the superfield KMS condition \equ{skms} is of bosonic type, 
since chiral and antichiral superfields are bosonic objects.
This condition can be proven directly~\cite{dl} at the thermal superfield 
level. Also, one easily verifies~\cite{dl} that the super-KMS condition yields 
the component KMS conditions \equ{c1}, \equ{c3} and \equ{c2}. The 
antiperiodicity \equ{antiper}, which captures the essence 
of thermal superspace, is an essential ingredient of these proofs. 

Our thermal superfield expansion \equ{tch} can easily be
seen to be consistent also from the point of view of the fields'
boundary conditions.
Thermal chiral superfields being bosonic objects, they
obey in thermal superspace a {\it periodic superfield 
boundary condition},
$
\widehat\phi[\widehat y_{(t)},\th(t)]
=\widehat\phi[\widehat y_{(t+i\b)},\th(t+i\b)]
$,
which writes, when cast in the variables $(x,\th,\tbh)$, $x=(t;\xx)$,
\be
\widehat\phi[t;\xx,\th(t),\tbh(t)]=
\widehat\phi[t+i\b;\xx,\th(t+i\b),\tbh(t+i\b)]\,.
\label{ssbc}
\ee
Indeed, expanding both sides in thermal superspace along
$$
\widehat\phi[x,\th,\tbh]=
z +\sqrt{2}\th\psi-\th\th f -i(\th\sigma^\m\tbh)\pa_\m z\nn\\
+{i\over\sqrt{2}}\th\th([\pa_\m\psi]\sigma^\m\tbh)
-{1\over 4} \th\th\tbh\tbh\Box z
\,,
$$
we get from \equ{ssbc}~: {\it (i)} for the thermal scalar field $z$ the
periodic b.c.
$z(t;\xx)=z(t+i\b;\xx)$, {\it (ii)}
for the thermal fermion $\psi$, upon replacing
$\th(t+i\b)=-\th(t)$, the antiperiodic b.c.
$\psi(t;\xx)=-\psi(t+i\b;\xx)$,
and {\it (iii)} for the thermal scalar field $f$, due to
$\th(t+i\b)\th(t+i\b)=\th(t)\th(t)$, the periodic b.c.
$f(t;\xx)=f(t+i\b;\xx)$.
%
%

\vspace{7mm}
\noindent{\large\bf 3 \ Thermal Covariant Derivatives}
\vspace{4mm}

%
Deriving supersymmetry covariant derivatives and supercharges  on
thermal superspace can be done simply by 
evaluating the effect of changing  variables
from  $T\!=\!0$ superspace $[x^\mu,\t,\tb]$ to thermal 
superspace $[x'^\mu,\t',\tb']=[x^\mu,\th(t),\tbh(t)]$. 
The partial derivatives with respect to
$\xx$, $\t$ and $\tb$ transform trivially, while
$\pa_t\rightarrow(\pa_t t')\,\pa_{t'} + (\pa_t\th^\a)\,\pah_\a + 
(\pa_t\tbh^\ad)\,\pabh_\ad$. Setting $\pa_t t'=1$, we define the 
{\it thermal space-time derivative} $\pah_\mu$ 
-- the thermal covariantization of $\pa_\m$ -- as
\be
\pah_\m=
( \pa_t - \Delta \ \ ; \ \vec\pa )\,,\qquad
\Delta \equiv (\pa_t\th^\a)\,\pah_\a + (\pa_t{\tbh^\ad})\,\pabh_\ad\,,
\label{pamch}
\ee
which obeys $\pah_\m\th(t)=\pah_\m\tbh(t)=0$.
The {\it thermal covariant derivatives} are then defined as~:
$$
\widehat D_\a = \pah_\a -i\,\s^\m_\aad\tbh^\ad \widehat\pa_\m\,,
\qquad
\widehat{\bar D}_\ad = \pabh_\ad -i\, \th^\a\s^\m_\aad \widehat\pa_\m\,,
$$
with $\pah_\a\equiv\pa/\pa{\th^\a}$, $\pabh_\ad\equiv\pa/\pa{\tbh^\ad}$. 
$\widehat D_\a$, $\widehat{\bar D}_\ad$ close on thermal translations and 
obey the $T\!=\!0$ ACRs 
$\{ \widehat D_\alpha , \widehat{\bar D}_\ad \} =
-2i\sigma^\mu_{\alpha\ad}\widehat\partial_\mu$,
$\{\widehat D_\a , \widehat D_\b \}
= \{ \widehat{\bar D}_\ad , \widehat{\bar D}_\bd \} =0$.
Furthermore, the thermal covariant derivatives provide a {\it definition} 
of the thermal chiral and antichiral superfields as the solution to 
$\widehat{\bar D}_\ad\ \widehat\phi = 0$, 
$\widehat D_\a\ \widehat{\bar\phi}= 0$.
%
%

\vspace{7mm}
\noindent{\large\bf 4 \ Thermal Supercharges and the 
Thermal Supersymmetry\\ \phantom{\large\bf 4 } Algebra}
\vspace{3mm}

%
The thermal supercharges are constructed using the same procedure as 
for the thermal covariant derivatives, that is, we replace
$\t$, $\tb$ by $\th$, $\tbh$, and $\pa_\m$, $\pa_\a$, $\pab_\ad$
by $\widehat\pa_\m$ [eq. \equ{pamch}], $\pah_\a$ and $\pabh_\ad$.
This yields the {\it thermal supercharges}~:
$$
\ba{rclrcl}
\qh_\a
\=
-i\pah_\a + \s^\m \tbh^\ad\pa_\m - \s^0_\aad\tbh^\ad
( \pa_t{\th^\g}\,\pah_\g + \pa_t{\tbh^\gd}\,\pabh_\gd)
&
\!\!\= -i\,\widehat\pa_\a + \s^\m_\aad\tbh^\ad \widehat\pa_\m
\,,\\[1mm]   
\qbh_\ad 
\= 
\phantom{-} i\pah_\ad - \th^\a\s^\m_\aad \pa_\m + \th^\a\s^0_\aad
( \pa_t{\th^\g}\,\pah_\g +\pa_t{\tbh^\gd}\,\pabh_\gd)
&
\!\!\= \phantom{-} i\,\widehat\pa_\ad - \th^\a\s^\m_\aad \widehat\pa_\m
\,,
\ea
$$
for which, as at $T\!=\!0$,
$\{ \qh , \dh \} =
\{ \qbh , \dh\} =
\{ \qh , \dbh \} =
\{ \qbh , \dbh \} = 0$.
For the super-Poincar\'e algebra at finite 
temperature\footnote{The thermal translation and 
Lorentz generators are defined similarly~\cite{dl}.}, one gets 
{\it the same structure as at $T\!=\!0$, provided one has appropriately 
covariantized all the generators with respect to thermal 
superspace}. In particular\footnote{The thermal time translation operator 
$\widehat P^0=-i\widehat\pa^0$ can be interpreted as a central 
charge of the subalgebra one obtains upon removing  the thermal 
Lorentz boosts~\cite{dl}.} \cite{dl}
\be
\{ \qh_\a , \qbh_\bd\} 
\ =\ -2\, \s^\m_{\a\bd} \widehat P_\m
\ =\ -2\, (\s^0_{\a\bd} \widehat P_0 - \s^i_{\a\bd}P_i )\,,
\qquad \widehat P_i=P_i\,.
\ee
However, the fact that the algebra is preserved at finite temperature does 
{\it not} mean that supersymmetry is thermally unbroken. 
This can be understood as follows.
Both the thermal fields' b.c.'s and the KMS
conditions carry information that is of 
{\it global} character,
in the sense that it relates the values of the field at distant
regions in space-time, along the time direction. As a consequence, the 
thermal superalgebra, which is a {\it local} structure, is insensitive 
to such global conditions and preserves its structure at finite temperature.
In particular, the antiperiodicity conditions on the thermal Grassmann 
coordinates, eq. \equ{antiper}, have no influence on the algebra. It is 
only the {\it local} statement that the superspace Grassmann variables
should be allowed to depend on time which makes it
necessary to covariantize the algebra generators.

To analyze thermal supersymmetry breaking, we must invetigate systems of 
thermal fields. As thermal bosons and fermions are distinguished by their 
global boundary and KMS conditions, we shall see thermal supersymmetry 
breaking in doing so.
{\it E.g.}, expanding {\it \`a la Matsubara} the $d=4$, $T\!=\!0$ 
Wess-Zumino action, we get a $d=3$ Euclidean action for the thermal 
modes in which the mass degeneracy is seen to be broken~\cite{dl}. 
%
%

\vspace{4mm}
\noindent{\large\bf 5 \ Field Realizations}
\vspace{2mm}

%
Thermal supersymmetry breaking should also be seen at the level of the 
{\it realizations} of the thermal supersymmetry algebra.
The next question we ask, therefore, is how thermal supersymmetry transforms 
the components of thermal superfields. This means translating into component 
language the thermal supersymmetry transformation
$\widehat\delta\widehat\phi= i
(\widehat\epsilon^\a\qh_\a +\widehat{\bar\epsilon}_\ad\qbh^\ad)
\widehat\phi$.
We may either derive straightforwardly the component transformations~\cite{dl} 
and note the perfect analogy to the $T\!=\!0$ case, or invoke the 
argument~\cite{ggs} that the thermal supersymmetry parameters should 
become time-dependent and antiperiodic at finite temperature. In both cases, 
we get
\vspace{-1.5mm}
\be
\ba{rcl}
\widehat\delta z\=
\sqrt{2}\widehat\epsilon^\a \psi_\a \,,
\qquad\qquad\qquad\qquad\qquad
\widehat\delta f
\ =\ -i\sqrt{2} (\s^\m \widehat{\bar\epsilon})^\a (\pa_\m \psi_\a)\,,
\\[1mm]
\widehat\delta \psi_\a\=
-\sqrt{2}\widehat\epsilon_\a f -i\sqrt{2}
(\s^\m \widehat{\bar\epsilon})_\a (\pa_\m z)\, ,
\ea
\label{tt2}
\ee
where {\it the unique difference with
the case of zero temperature is the appearance of the thermal
(time-dependent and antiperiodic) spinorial parameter $\widehat\epsilon$, 
$\widehat{\bar\epsilon}$}, with
$\widehat\epsilon(t+i\b)=-\widehat\epsilon(t)$,
$\widehat{\bar\epsilon}(t+i\b)=-\widehat{\bar\epsilon}(t)$,
in place of the constant spinorial parameter $\epsilon$, $\bar\epsilon$ of 
$T\!=\!0$ supersymmetry. Eqs. \equ{tt2} 
can be translated into transformations 
of the $d\!=\!3$  Matsubara modes~\cite{dl}, upon developing thermally the 
$t$-dependent parameters $\widehat\epsilon$, $\widehat{\bar\epsilon}$.
Neither the ($T\!=\!0$) Wess-Zumino kinetic action nor the mass action are
invariant under the thermal supersymmetry transformations.
At the level of the $d\!=\!3$ thermal modes, the variation of the total 
action is seen to be proportional to the 
fermionic Matsubara frequency $\o_n^F \sim T$.
In the $T\rightarrow 0$ limit, one expects supersymmetry
to be restored. The variations of the mass and kinetic actions
indeed vanish separately in that limit.
%
%
\vspace{-4mm}
{}
%
\end{document}